\documentclass[11pt,a4paper]{article}

\usepackage{graphicx}
\usepackage{amsmath}
\usepackage{amssymb}
\usepackage{url}
\usepackage{microtype}
\usepackage{setspace}
\usepackage{caption}
\usepackage[hidelinks]{hyperref}
\usepackage{wrapfig}
\usepackage{titlesec}
\usepackage{cite}
\usepackage[top=1.0in, bottom=1.0in, left=1.25in, right=1.25in]{geometry}
\usepackage[utf8]{inputenc}
\usepackage{xcolor}
\usepackage{wrapfig}
\usepackage{booktabs}

\newcommand{\iec}{i.\,e.,\ }

\newcommand{\egc}{e.\,g.,\ }

\newcommand{\Fig}[1]{Figure~\ref{#1}}

\newcommand{\D}{\mathcal{D}}
\newcommand{\Qu}{\mathcal{Q}}

\def\N{{\mathbb N}}

\def\R{{\mathbb R}}

\newcommand{\chroma}[1]{\mathrm{#1}}
\newcommand{\chromaSharp}[1]{\mathrm{#1}^{\sharp}}

\hypersetup{
   pdfauthor = {Meinard M\"uller, Andreas Arzt, Stefan Balke, Matthias Dorfer, Gerhard Widmer},
   pdftitle = {Cross-Modal Music Retrieval and Applications}
}

\def\maintextlinespacing{1.59}

\def\appendixlinespacing{1.59}

\titlespacing*{\section}
{0pt}{1.2ex}{0.9ex}
\titlespacing*{\subsection}
{0pt}{1.2ex}{0.9ex}

\def\N{{\mathbb N}}

\def\R{{\mathbb R}}

\title{\vspace{-6ex} Cross-Modal Music Retrieval and Applications\thanks{\textcopyright 2019 IEEE. Personal use of this material is permitted. Permission from IEEE must be obtained for all other uses, in any current or future media, including reprinting/republishing this material for advertising or promotional purposes, creating new collective works, for resale or redistribution to servers or lists, or reuse of any copyrighted component of this work in other works.The published version is available online: \url{https://doi.org/10.1109/MSP.2018.2868887}}}
\author{\emph{Meinard M{\"u}ller$^{\star}$, Andreas Arzt$^{\dagger}$, Stefan Balke$^{\star}$, Matthias Dorfer$^{\dagger}$, Gerhard Widmer$^{\dagger\ddagger}$}\vspace{-0.5ex} \\ \\ \small $^{\star}$International Audio Laboratories Erlangen, Erlangen, Germany \\ \small $^{\dagger}$Johannes Kepler University, Linz, Austria\\  \small $^{\ddagger}$Austrian Research Institute for Artificial Intelligence (OFAI), Vienna, Austria\vspace{-2ex}}
\date{}

\begin{document}


\maketitle

\linespread{\maintextlinespacing}
\selectfont


\section*{Introduction}
\label{sec:Introduction}

There has been a rapid growth of digitally available music data including audio recordings, digitized images
of sheet music, album covers and liner notes, and video clips. This huge amount of data calls for retrieval strategies that allow users to explore large music collections in a convenient way.
More precisely, there is a need for \emph{cross-modal} retrieval algorithms that, given a query in one modality (\egc a short audio excerpt), find corresponding information and entities in other modalities (\egc name of piece and sheet music).
This goes beyond \emph{exact audio identificiation} (and subsequent retrieval of meta-information)
as performed by commercial applications like Shazam~\cite{Wang03_Shazam_ISMIR}.

In this paper, we review several cross-modal retrieval scenarios, with a particular focus on sheet music (visual domain) and audio (acoustic domain).
First, we discuss a traditional approach where the sheet music and audio representations are converted into common \emph{mid-level feature representations that capture musical properties related to pitches and harmony.}
The resulting feature sequences can then be compared using standard alignment algorithms~\cite{KurthMFCC07_AutomatedSynchronization_ISMIR,Mueller15_FMP_SPRINGER}.
Second, we review an approach based on \emph{symbolic fingerprinting techniques}. Originally, audio fingerprinting refers to a procedure that allows for a robust identification of exact replicas of audio recordings~\cite{CanoBKH05_FingerprintingReview_VLSI}.
In our cross-modal scenario,
we discuss tempo- and transposition-invariant symbolic fingerprinting methods based on note parameters extracted  via audio transcription techniques~\cite{ArztBW12_SymbolicFingerprint_ISMIR,ArztWS14_TempoTranspInvariantIdent_ISMIR}.
Third, employing \emph{deep learning methods}, we describe an end-to-end cross-modal retrieval strategy that works without needing manually crafted feature representations~\cite{DorferAW17_ScoreIdentification_ISMIR}. Given snippets of sheet music (in the form of pixel images) and corresponding audio excerpts (in the form of spectrograms), a neural network learns a joint embedding space, on which cross-modal retrieval can be performed using simple distance measures and nearest-neighborhood search.

Using these three approaches as illustrative examples, the primary objective of this paper is to discuss principles and challenges encountered in general music processing, such as designing musically motivated features and similarity measures to cope with semantic data variability.
Furthermore, to illustrate the potential of cross-modal retrieval techniques, we describe some navigation and browsing applications including a prototype system called the \emph{Piano Music Companion}, while indicating future research directions.

\section*{Music Representations}
\label{sec:modal}

\begin{figure}[t]
  \centering
  \includegraphics[width=12cm]{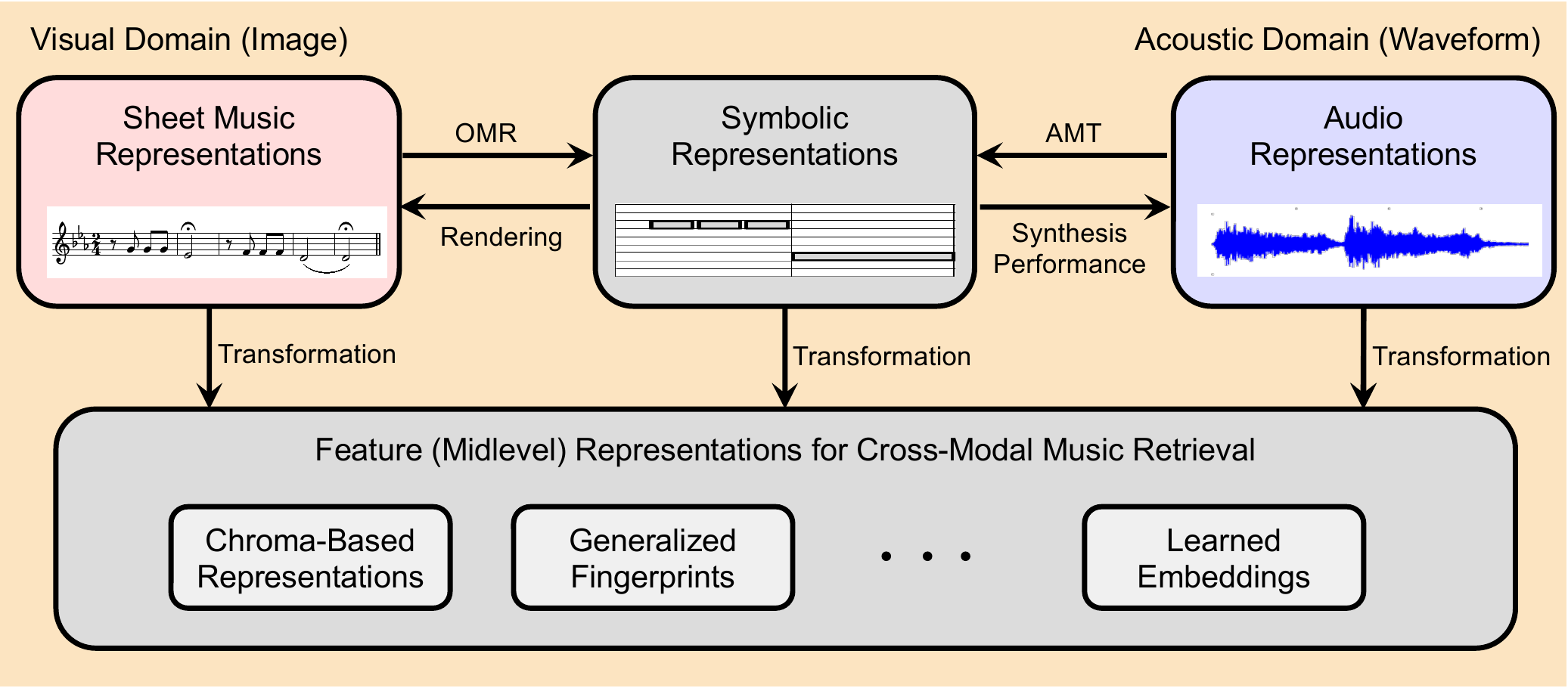}
  \caption{\small The different representations for music data and data transformations
  relevant for cross-modal music retrieval.}
\label{fig:MusicRep}
\end{figure}

Before we delve into the various cross-modal retrieval approaches, we first introduce some basic notions, following~\cite[Chapter 1]{Mueller15_FMP_SPRINGER}. As indicated by~\Fig{fig:MusicRep}, music can be represented in many different ways and formats. For example, a composer may write a composition in the form of a musical score, where musical symbols are used to visually encode which notes are to be played, and how. The printed form of a musical score is also referred to as \emph{sheet music}. The original medium of this representation is paper, although it is now also accessible on computer screens in the form of digital images.
In electronic instruments and computers, music can be communicated by means of standard
protocols (such as the widely used MIDI\footnote{\url{https://www.midi.org/}} protocol), where event messages specify note pitches, note intensities (velocities), and other parameters to generate the intended sounds. Often, the term \emph{symbolic} is used to refer to any data format that explicitly represents musical entities. The musical entities may range from timed note events, as is the case in MIDI files, to graphical shapes with attached musical meaning, as is the case in music engraving systems.
In contrast to such symbolic representations, the musical events are not given explicitly in \emph{audio representations} such as WAV or MP3 files. These encode acoustic waves that are generated when, \egc playing an instrument, and travel from the sound sources to the human ear as air pressure oscillations.

At this point it is important to note that each of these representations reflects certain aspects of a musical entity, but no single representation encompasses all properties. For example, rather than giving strict specifications, a musical score only serves as a guide for performing a piece of music, leaving room for different interpretations. Reading the instructions in the score, a musician shapes the music by varying the tempo, dynamics, articulation, and other parameters, thus creating a personal interpretation of the piece.
Furthermore, while sheet music visually encodes the musical notes, such information is hidden in an audio recording, which is basically a time series of samples. In summary, even if they refer to the same piece of music, there may be a significant gap---technically as well as semantically---between different representations such as sheet music and audio.

The boundaries between the various music representations are not sharp. As illustrated by \Fig{fig:MusicRep}, symbolic representations---depending on their specific format and intended application---may be closer to sheet music or audio representations.
For example, symbolic representations such as MusicXML\footnote{\url{https://www.musicxml.com/}} are used for \emph{rendering} sheet music, where the shape of the note objects and their arrangement on a page are determined. Optical music recognition (OMR) can be seen as the inverse process with the objective to transform sheet music into a symbolic representation. Furthermore, symbolic representations such as MIDI are used for \emph{synthesizing} audio, where the note objects are transformed into musical tones and real sounds. The inverse process is known as \emph{automatic music transcription} (AMT) and aims at extracting note events, key signature, time signature, instrumentation, and other score parameters from a given music recording~\cite{Mueller15_FMP_SPRINGER}.
Both transformations, OMR as well as AMT, are far from straightforward. For example, correctly recognizing and interpreting the meaning of all the musical symbols in complex sheet music is easy for a trained human, but hard for a computer. Even though current OMR software is reported to yield highly accurate results, manual postprocessing is necessary to obtain a high-quality symbolic representation~\cite{Byrd2015_OMR_JNMR}. Similarly, converting a music recording into a note-based representation is a largely unsolved problem---in particular, for multi-voiced music involving different instruments~\cite{BenetosDGKK13_MusicTranscription_JIIS}.

For relating different types of data (\egc sheet music and audio data) to each other, traditional methods are often based on mid-level representations that exploit specific domain knowledge.
As an important example, we first consider \emph{mid-level representations} that capture musical properties related to pitches and harmony.
We then discuss symbolic fingerprints that are based on note-level descriptors. Both of these approaches require expert knowledge in the transformation process.
As an alternative, we present an \emph{end-to-end learning} approach based on deep neural networks, where the idea is to circumvent the explicit definition of a mid-level representation. In the following sections, we address benefits and limitations of these conceptually different approaches in the context of cross-modal music retrieval.

\section*{Chroma-Based Approach}
\label{sec:retrieval:DP}

\begin{figure}[t]
  \centering
  \includegraphics[width=12cm]{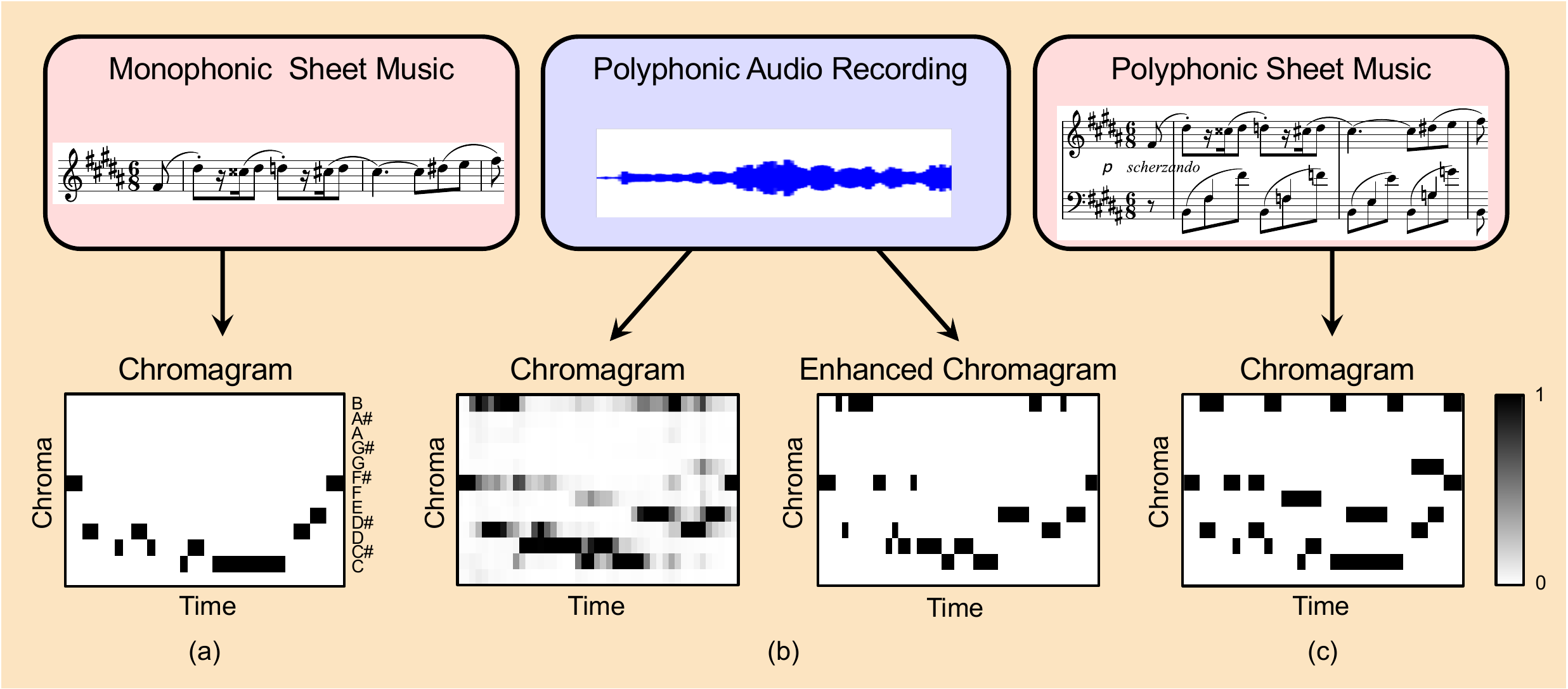}
  \caption{\small Some chromagrams obtained from (a) monophonic and (c) polyphonic sheet music and (c) polyphonic audio representations for the beginning of Frédéric Chopin's Nocturne in B Major, Op. 9, No. 3.}
\label{fig:Chroma}
\end{figure}

To make music data algorithmically accessible, traditional music processing tries to extract suitable features that capture relevant key aspects while suppressing irrelevant details. For music-related retrieval and analysis tasks, \emph{chroma features} have turned out to be a powerful mid-level representation~\cite{Gomez06_PhD,Mueller15_FMP_SPRINGER}.

Due to their central importance in music processing, we give a short introduction to the basics of chroma features following \cite[Chapter 1]{Mueller15_FMP_SPRINGER}. Recall that playing a note on an instrument results in a (more or less) periodic sound of a certain fundamental frequency. This fundamental frequency is closely related to what is called the \emph{pitch} of a note. This notion allows us to order pitched sounds from ``lower'' to ``higher''---similarly to the keys of a piano keyboard ordered from left to right. Two notes with fundamental frequencies in a ratio equal to any power of two (\egc half, twice, or four times) are perceived as very similar (or musically/harmonically equivalent, in some sense). This observation leads to the fundamental notion of an \emph{octave}, which is defined as the interval between one musical note and another with half or double its fundamental frequency. In Western music, the ``space'' within one octave is generally subdivided into twelve scale steps with fundamental frequencies equally spaced on a logarithmic frequency axis, resulting in what is known as the \emph{twelve-tone equal-tempered scale}. In this scale, each pitch can be separated into two components, which are referred to as \emph{tone height} (or octave number) and \emph{chroma} (or pitch spelling attribute denoted by $\chroma{C}$, $\chromaSharp{C}$, $\chroma{D}$, $\ldots$, $\chroma{B}$ in Western music notation).

Chroma features rely on this perception of octave equivalence and map absolute pitch into twelve octave-independent pitch classes, where a pitch class consists of all pitches that share the same chroma.
Thus, a chroma feature is represented by a $12$-dimensional vector $x=(x(1),\ldots, x(12))^\top$, where $x(1)$ corresponds to chroma $\chroma{C}$, $x(2)$ to $\chromaSharp{C}$, and so on. In the feature extraction step, a given audio signal is converted into a sequence of chroma vectors (also called \emph{chromagram}), where each vector expresses how the short-time energy of the signal is spread over the twelve chroma bands. A chromagram closely correlates to the melodic and harmonic progression of the music, while exhibiting a high degree of robustness to variations in instrumentation and dynamics.

There are many ways for computing chroma-based features from audio recordings, \egc using short-time
Fourier transforms (STFT) in combination with binning strategies~\cite{Gomez06_PhD}
or by employing suitable multirate filter banks~\cite{Mueller07_InformationRetrieval_SPRINGER}. Furthermore, the properties of chroma features can be significantly changed by introducing suitable pre- and post-processing steps modifying spectral, temporal, and dynamical aspects. As an example, \Fig{fig:Chroma} (center part) shows two different chromagram variants extracted from a piano audio recording. While the first one is a traditional chromagram, the second version is enhanced such that certain important frequencies that relate to melody notes as specified by the upper staff, are emphasized---which can be important,\egc~for melody-based retrieval.
When given a symbolic music representation (such as MIDI or MusicXML files), it is straightforward to derive chromagrams from the explicitly encoded note parameters (pitches, note onsets, note durations). \Fig{fig:Chroma} shows a symbolic chromagram obtained from a monophonic (left part) and polpyhonic (right part) sheet music representation.
While symbolic chromagrams are based on ``pure'' note information, audio-based chromagrams tend to be ``noisy'', reflecting the full range of the signal's acoustic properties (including partials, transients, room acoustics).
Still, as also demonstrated by \Fig{fig:Chroma}, chroma features mainly capture melodic and harmonic properties and are suited to serve as a mid-level feature representation for comparing and relating acoustical and symbolic music.

\begin{figure}[t]
  \centering
  \includegraphics[width=12cm]{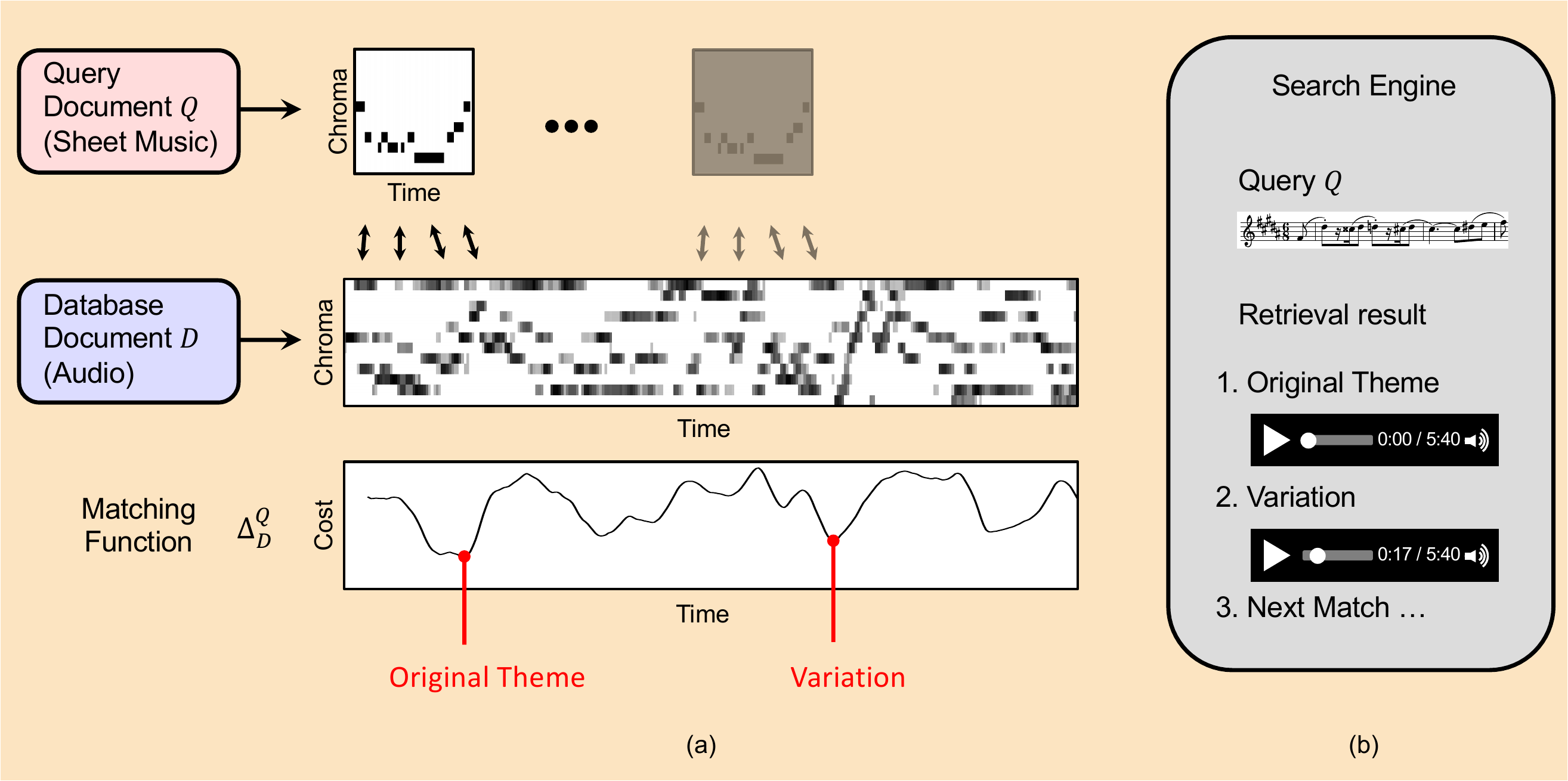}
  \caption{\small An (a) illustration of the matching procedure with chroma-based representations. (b) The costs of the local minima yield a natural ranking of the retrieved documents and their relevant sections, which are shown in the form of a ranked list.}
\label{fig:RetrievalChroma}
\end{figure}

To demonstrate the applicability and potential of chroma-based features, we consider a cross-modal retrieval scenario motivated by Barlow and Morgenstern's book \emph{A Dictionary of Musical Themes} published in 1949~\cite{BarlowM75_MusicalThemes_BOOK}. This book contains about 10,000 musical themes of well-known instrumental pieces from the corpus of Western Classical music. These monophonic themes (usually four bars long) are typically the most memorable parts of a piece of music.
This motivates the retrieval scenario as considered in~\cite{BalkePM15_MatchingMusicalThemes_ICASSP,BalkeALM16_BarlowRetrieval_ICASSP}, where the objective is to retrieve all audio recordings from a music collection that contain a specified musical theme.
More formally, let $\Qu$ be the collection of musical themes, where each element $Q\in\Qu$ is regarded as a \emph{query}. Furthermore, let $\D$ be a set of audio recordings, which we regard as a database collection consisting of \emph{documents} $D\in\D$. Given a query $Q\in\Qu$, the retrieval task is to identify the
semantically corresponding documents $D\in\D$.
One approach, as illustrated by \Fig{fig:RetrievalChroma}, is to first transform a query $Q$ (possibly using OMR as an intermediate step) and each of the documents $D$ into chromagrams. Based on these mid-level representations, one computes a matching function $\Delta_D^Q$ by locally comparing the query chromagram to the audio chromagram using a subsequence variant of dynamic time warping (DTW), see~\cite[Chapter 4]{Mueller07_InformationRetrieval_SPRINGER}. For each position of the audio recording $D$, such a matching function indicates the local cost of aligning the query chromagram with a segment ending at that position of the audio chromagram. In other words, each local minimum of $\Delta_D^Q$ that is close to the value zero points to a location where the query (musical theme) is similar to a local segment of the document (audio recording). Thus, for a given query, the retrieval task can be solved by computing matching curves for all documents and screening for local minima that are below a certain threshold in these curves. The costs of the local minima yield a natural ranking of the retrieved documents and their relevant sections, which can then be presented in the form of a ranked list, see \Fig{fig:RetrievalChroma} (right side).

As detailed in~\cite{BalkePM15_MatchingMusicalThemes_ICASSP,BalkeALM16_BarlowRetrieval_ICASSP}, there are various challenges that need to be addressed, including tempo deviations, OMR extraction errors, musical tunings, key transpositions, and differences in the degree of polyphony between the symbolic query and the audio recordings.
For some of these challenges, there already exist reliable compensation strategies. For example, key transpositions are simulated by a cyclic shift of the query's chromagram, or local and global tempo deviations are compensated by using sequence alignment techniques such as DTW. Handling differences in the degree of polyphony is still subject to ongoing research. One strategy to bridge the ``polyphony gap'' is to first extract the predominant melody of the audio recording using harmonic summation~\cite{SalamonSG13_Retrieval_IJMRI} and source-filter models~\cite{BoschBSG16_MelodyExtraction_ISMIR}. From the resulting salience representations, enhanced audio chromagrams that better match the monophonic theme may be derived (see \Fig{fig:Chroma} for an illustration).

Obviously, computing matching curves for each database document results in a retrieval procedure that does not scale to large music collections. Indexing techniques based on short audio excerpts (so called~\emph{audio shingles})
can help speed up the retrieval procedure~\cite{CaseyRS08_MinimumDistances_IEEE-TASLP,GroscheM12_RetrievalShingles_ICASSP}. In the next section, we discuss an alternative approach that is based on symbolic fingerprints and permits extremely efficient retrieval.

\section*{Symbolic Fingerprinting Approach}
\label{sec:retrieval:FP}

We have seen that chroma features are a very convenient mid-level representation for comparing music data of different modalities. One main benefit is that both symbolic and audio data can be easily converted into chromagrams. Furthermore, capturing only the coarse harmonic/melodic progression, chromagrams are highly robust to musical and acoustic variations.
However, the reduction onto the chroma level also leads to a loss of valuable information that may be contained in the input data (such as accurate timing and pitch parameters as encoded by sheet music).
As a consequence, chroma-based retrieval strategies often become problematic for short input sequences (\egc covering only a couple of notes). Furthermore, reducing pitch information to the twelve chroma bands renders the comparison of monophonic and polyphonic versions difficult.
An alternative to using chroma-based features is to exploit the high specificity of note parameters and of resulting time--pitch patterns of occurring notes.
To this end, both the visual and acoustic data need to be transformed into the symbolic music domain. In the following, we discuss such an approach based on symbolic fingerprints and highlight the resulting benefits and limitations.

Traditionally, in music processing, audio fingerprinting refers to methods for identifying \emph{exact replicas} of audio recordings, which are possibly distorted in some way (\egc compression artifacts or background noise). For this problem, also known as \emph{audio identification}, powerful algorithms exist and are in everyday use in commercial applications (see, \egc~\cite{CanoBKH05_FingerprintingReview_VLSI,Mueller15_FMP_SPRINGER,SonnleitnerW16_QuadFingerp_TASLP,Wang03_Shazam_ISMIR}).
In the identification process, the audio material is compared by means of so-called \emph{audio fingerprints}, which are compact and discriminative audio features. There are many different ways of designing and computing audio fingerprints, and the suitability of a specific type of fingerprint very much depends on the requirements imposed by the intended application. For example, in the pioneering work by Wang~\cite{Wang03_Shazam_ISMIR}, a fingerprinting approach is described that operates on spectral peaks extracted from a time--frequency representation.
Recent work such as~\cite{SixL14_PanakoAcousFP_ISMIR,SonnleitnerW16_QuadFingerp_TASLP} has focused on making fingerprinting algorithms more robust to transformation in the time (playback speed of the audio) and the frequency scale (transpositions).
Classical fingerprinting approaches, combined with indexing techniques, allow for a very efficient (scalable to huge fingerprint datasets) and effective (high precision even for short queries) identification of audio material.
However, being based on audio-specific spectro-temporal patterns, these techniques are not suited for handling music-specific variations as required for cross-modal music retrieval or related tasks such as cover song retrieval~\cite{Mueller15_FMP_SPRINGER,SerraGH10_coversong_BOOKCHAP}.

\begin{figure}[t]
  \centering
  \includegraphics[width=4.5cm]{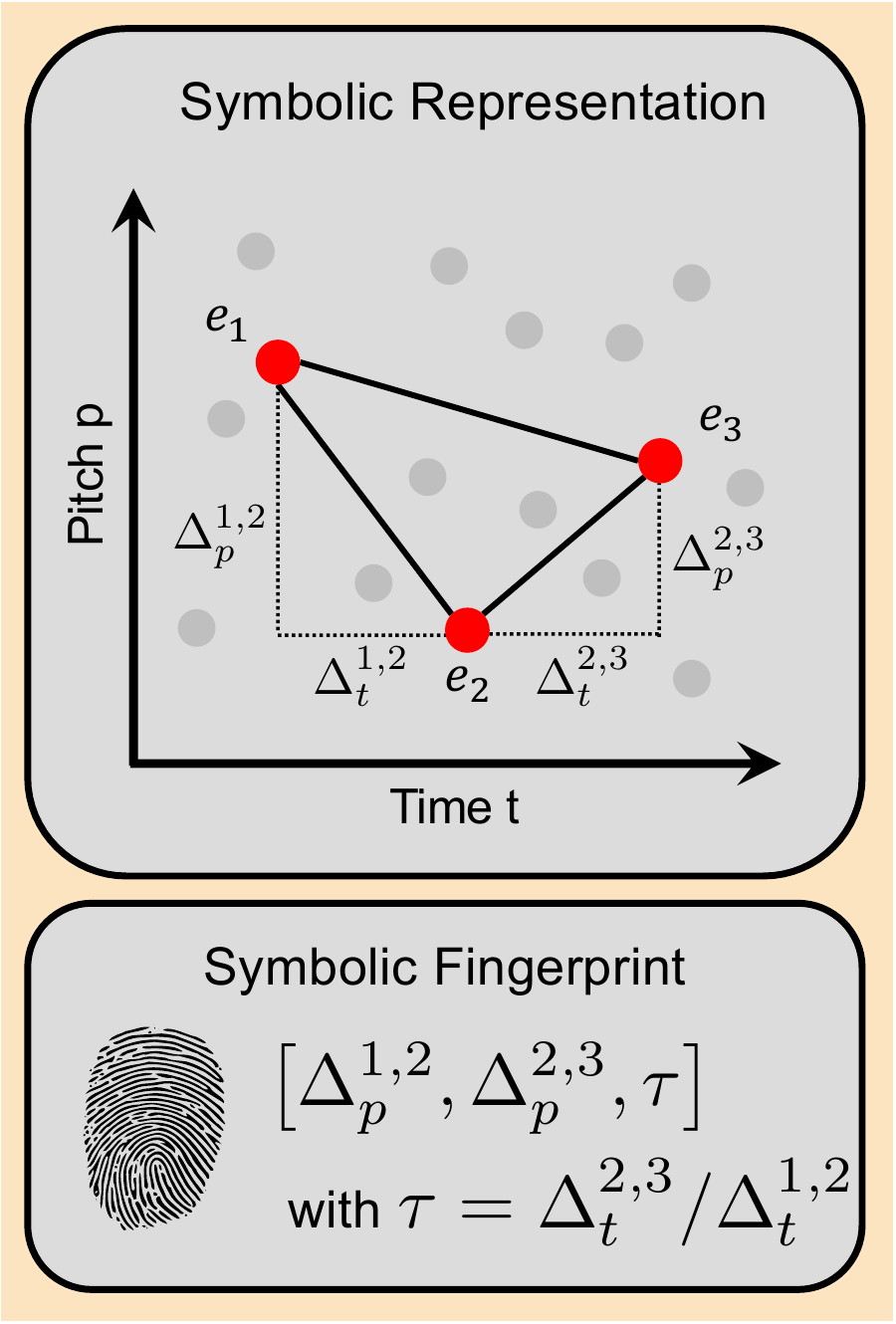}
  \caption{\small An illustration of symbolic fingerprints.}
\label{fig:FingerPrinting}
\end{figure}

Inspired by classical fingerprinting techniques, Arzt et al.~\cite{ArztBW12_SymbolicFingerprint_ISMIR,ArztWS14_TempoTranspInvariantIdent_ISMIR} introduced a symbolic fingerprinting approach, which not only allows for the identification of exact replicas of recordings, but also for fast retrieval of \emph{different versions} of the same piece of music including differently performed audio recordings and score representations.
In the following, we summarize the main idea of this approach.
For the moment, we start with a symbolic music representation where all note events are encoded explicitly.
As illustrated by \Fig{fig:FingerPrinting}, we assume that each note event $e=(t,p)$ is specified by an onset time $t$ and a pitch $p$. To obtain fingerprints, we consider triples consisting of three events $e_1=(t_1,p_1)$, $e_2=(t_2,p_2)$, and $e_3=(t_3,p_3)$ with $t_1<t_2<t_3$.
For each such triple, we define the time differences $\Delta_t^{1,2}:=t_2-t_1$ and $\Delta_t^{2,3}:=t_3-t_2$ as well as the pitch differences $\Delta_p^{1,2}:=p_2-p_1$ and $\Delta_p^{2,3}:=p_3-p_2$. Furthermore, we set $\tau:=\Delta_t^{2,3}/\Delta_t^{1,2}$. Finally, a \emph{symbolic fingerprint} is defined to be a list of the following numbers:
\begin{equation}
[\Delta_p^{1,2},\Delta_p^{2,3},\tau].
\end{equation}
Considering time and pitch relations in a \emph{relative} fashion, each fingerprint is invariant with regard to musical transpositions (pitch shifts) and tempo changes.
To obtain \emph{local} descriptors, fingerprints are computed only from note events within a certain neighborhood (typically a few seconds). This not only facilitates short query lengths, but also reduces the number of fingerprints to be stored in the database.
Also observe that since each individual fingerprint encodes relative timing information, we need to assume that the onset times of a triple are distinct. As a result, simultaneous note events (as occurring in a chord) may not be encoded by a single fingerprint. However, such co-occurring events can be captured by considering several fingerprints.
In summary, being discriminative yet compact descriptors of fixed length, such fingerprints have turned out to be suitable for indexing symbolically encoded music data.
%

\begin{figure}[t]
  \centering
  \includegraphics[width=11cm]{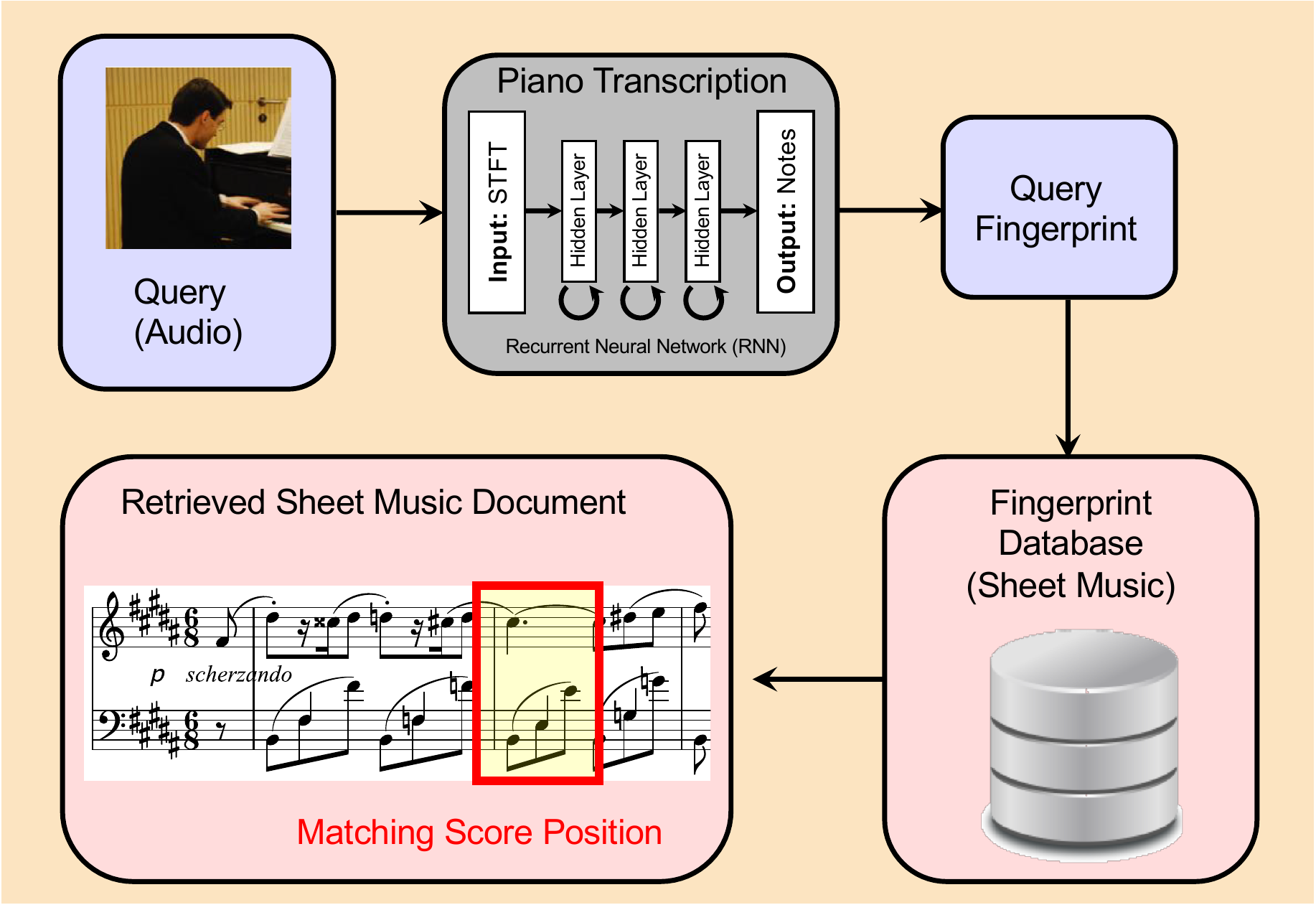}
  \caption{\small An illustration of cross-modal retrieval via piano transcription and symbolic fingerprinting. (Photo of Werner Goebl courtesy of Clemens Chmelar.)}
\label{fig:RetrievalFingerPrinting}
\end{figure}

We now discuss how the symbolic fingerprints can be used for cross-modal music retrieval. As a challenging example scenario, we consider a combined sheet-music identification and score following application tailored to piano music~\cite{ArztWS14_TempoTranspInvariantIdent_ISMIR}. Given a short excerpt of an audio recording (used as query), the task is to identify the underlying sheet music document as well as the exact score position (see \Fig{fig:RetrievalFingerPrinting}).
Accordingly, the database $\D$ for this task consists of sheet music representations of all pieces to be potentially identified. In a preprocessing step, all sheet music documents $D\in\D$ are first transformed into a suitable symbolic format (\egc by applying OMR or by extracting note parameters from a MusicXML file). From this encoding, symbolic fingerprints are extracted for each document by considering all possible triples of note events that obey certain constraints. For example, to avoid a combinatorial explosion, one typically imposes constraints in the form of minimum and maximum values for the time differences $\Delta_t^{1,2}$ and $\Delta_t^{2,3}$. The resulting fingerprints along with links to suitable metadata (\egc corresponding piece and sheet music positions) are stored in a fingerprint database that is equipped with efficient search structures based on indexing techniques.

Similarly, an incoming audio query is also transformed into a set of symbolic fingerprints. This, however, involves a non-trivial transcription step to convert the recording into a symbolic representation.
In general, automatic music transcription is still an unsolved
problem---in particular for polyphonic music recordings with many different instruments (\egc orchestral music),
see
\cite{BenetosDGKK13_MusicTranscription_JIIS,%
BoeckS12_TranscriptionRecurrentNetwork_ICASSP,%
SigtiaBD16_DNNPolyPianoTrans_TASLP}.
In the case of single-instrument polyphonic music (such as piano music), state-of-the-art algorithms provide reasonable, albeit far from perfect, transcriptions. In our scenario, we employ a recent transcription algorithm based on a recurrent neural network (RNN)~\cite{BoeckS12_TranscriptionRecurrentNetwork_ICASSP}.
The use of bidirectional hidden layers enables the system to better model the context of the notes, which exhibit a very characteristic envelope during their decay phase---in particular for piano music.
The network was trained on a collection of several hundred piano pieces recorded on various (virtual and real) pianos, see~\cite{BoeckS12_TranscriptionRecurrentNetwork_ICASSP} for further details.
To make the transcriber applicable also in online scenarios, instead of preprocessing the whole piece of audio at a time, the signal is split into blocks that consist of several subsequent frames centered around the current frame. Using such blocks (each covering roughly $210$~ms of audio) is a trade-off between keeping the system's ability to model the context of the notes and to keep the introduced delay at a minimum.
The network outputs a transcription of the audio query consisting of a list of note onsets and pitches, which can be further transformed into a set of audio fingerprints. Finally, the score fingerprint database is searched for subsets that approximately fit the query's set of audio fingerprints. The best matching subset in the database yields the sheet music document along with the score position, see \Fig{fig:RetrievalFingerPrinting}.

In contrast to chroma-based mid-level representations, symbolic fingerprints are compact, possess a high discriminative power, and are well suited for indexing techniques. As a result, these techniques scale well to large amounts of data in terms of memory requirements, accuracy, and efficiency. However, there is also a price to be paid. The necessary transcription from audio signals into the symbolic domain is a hard problem that is solvable well enough only for certain classes of music (\egc piano music recorded under reasonable acoustic conditions).
Even though a small proportion of the fingerprints extracted from the query may suffice to identify the correct piece, symbolic fingerprinting may fail if the input representation becomes too noisy.
For general music recordings including many instruments (\egc orchestra), there is still a long way to go; here one requires strategies that better adapt to the multitude of musical aspects including harmony, melody, rhythm, dynamics, and instrumentation.
In this context, recent advances in deep learning may help to make further progress in this area. In the subsequent section, we discuss such a deep learning approach that tries to learn sheet music and audio correspondences directly from raw input representations without the need for mid-level representations that explicitly exploit musical knowledge.

\section*{Deep Learning Approach}
\label{sec:retrieval:DL}

In the previous sections, we have seen two more traditional approaches for linking audio and sheet music data using musically informed mid-level representations---once using chroma features and once symbolic fingerprints.
Such representations not only require expert knowledge at the design stage, but are also problematic when relying on error-prone \mbox{(pre-)}processing steps such as automatic music transcription on the audio side or optical music recognition on the sheet music side.
As an alternative, we now present a methodology to \emph{directly} learn correspondences between audio data and sheet music images from a set of training observations, thus circumventing the explicit definition of a mid-level representation. This approach builds on the current success of artificial neural networks, nowadays often referred to as \emph{deep learning}, which have proven to be powerful tools for automatic feature learning~\cite{Schmidhuber15_DeepLearningOverview_NN}. Given snippets of sheet music images and corresponding audio excerpts, we introduce a cross-modal neural network that learns an embedding space in which both modalities are represented as low-dimensional vectors~\cite{DorferAW17_ScoreIdentification_ISMIR}. In this embedding space, cross-modal music retrieval can then be easily performed by using a simple similarity measure.

\begin{figure}[t]
  \centering
  \includegraphics[width=4.5cm]{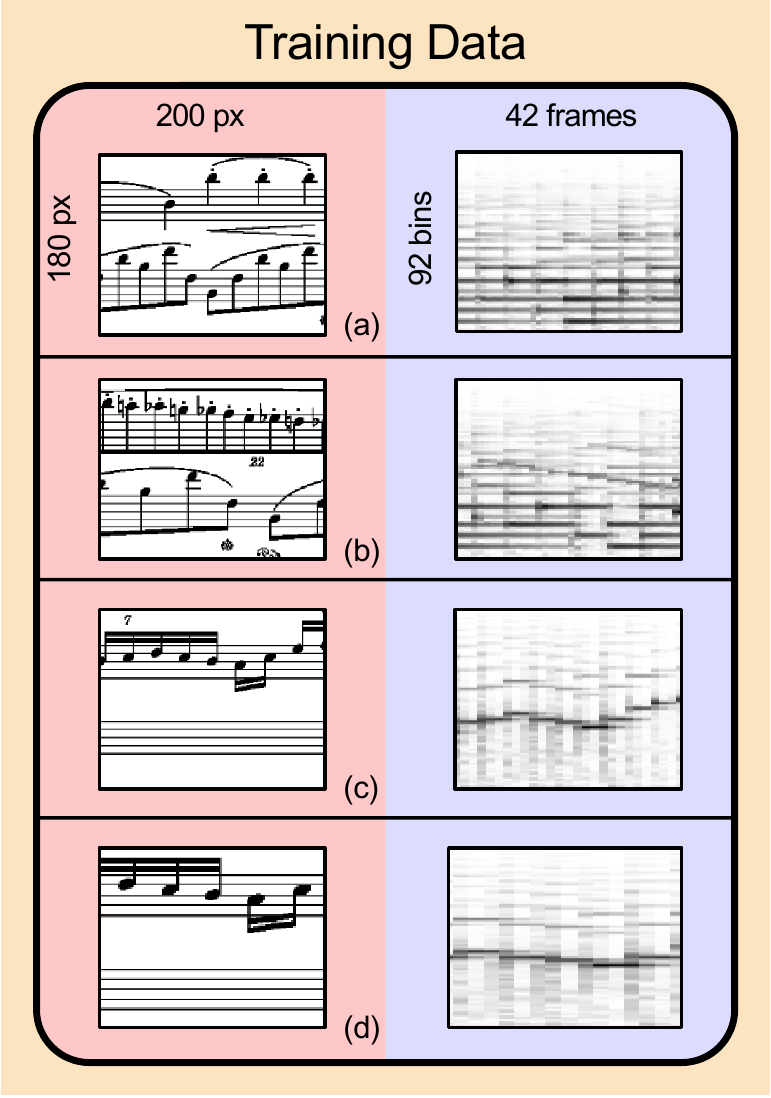}
  \caption{\small (a)--(d) Four training pairs, each consisting of a sheet music snippet and an audio excerpt. The pair in (d) was obtained from the pair in (c) by applying data augmentation techniques.}
\label{fig:DataDNN}
\end{figure}

The general principle of supervised feature learning is to learn latent representations in an end-to-end fashion from a set of raw training observations. Such approaches are not only generally applicable, but also have the advantage of automatically adapting the learned representations to the given problem. One limitation, however, is that supervised learning requires a sufficiently large set of training data to arrive at models that generalize well to unseen data.
In our scenario, we need training pairs that consist of sheet music snippets and corresponding audio excerpts. Typical examples as used in our system are shown in \Fig{fig:DataDNN}(a)--(d). Note that for creating such training pairs, we need to first establish correspondences between individual pixel locations of the note heads in a score and their respective counterparts (note onset events) in the corresponding audio recording. Establishing the correspondences can be done either in a manual annotation process or by relying on synthetic training data generated from digital sheet music formats such as \emph{Musescore}\footnote{\url{https://musescore.com/}} or \emph{Lilypond}\footnote{\url{http://lilypond.org/}}. Based on these relationships, one can generate corresponding snippets of sheet music images (in our case $180 \times 200$ pixels) and short excerpts of audio (in our case represented by log-frequency spectrograms with $92$ bins $\times$ $42$ frames). These are the pairs presented to the multi-modal network for training.
%
%
To improve the generalization ability of the resulting network, one can further apply \emph{data augmentation} techniques to (synthetically) increase the effective size of the training set and to better account for relevant data variability. In this setting, different transformations for sheet music augmentation (\egc image scaling and translation) and audio augmentation (\egc using different sound fonts and tempo scaling) are applied.
At this point, we emphasize that data augmentation is a crucial component for learning cross-modal representations that generalize to unseen music, especially when little data is available.
In this process, augmenting the dataset using data transformations is conceptually different and more promising than automatically generating random scores. First, rendering (synthesizing) sheet music typically results in images with strong regularities (\egc same scale or perfectly centered staff lines). By applying image transformations, these regularities are disturbed, thus making the embedding networks robust to small distortions as occurring in realistic scenarios, \egc images of printed sheet music scanned under different conditions and sheet music originating from different publishers using varying type settings. Second, note that music and hence also sheet music follows musical rules. Therefore, augmentation by adding randomly generated music may distort the inherent data distribution of ``realistic'' music, which may have a negative impact on embedding space learning.

\begin{figure}[t]
  \centering
  \includegraphics[width=11cm]{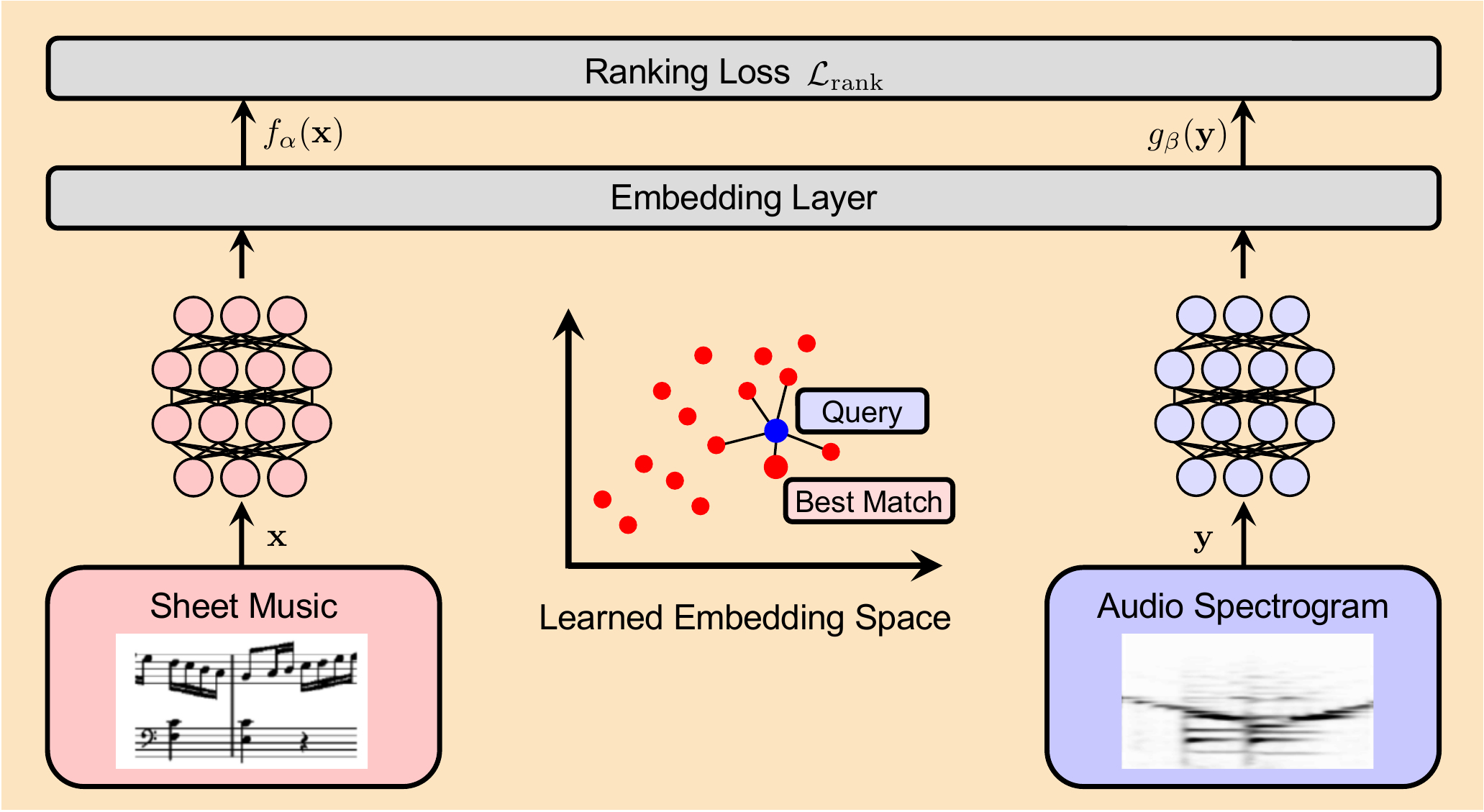}
  \caption{\small An illustration of the network used for learning a cross-modal embedding space.
   At application time, the learned functions $f_\alpha$ and $g_\beta$ are used to project the sheet music snippets and audio excerpts, respectively, to the joint embedding space.
}
\label{fig:RetrievalDNN}
\end{figure}

Based on such training pairs, the retrieval task is formulated as an embedding problem with the aim of learning a joint embedding space of the two different modalities~\cite{DorferAW17_ScoreIdentification_ISMIR}. This approach is inspired by a similar text--to--image retrieval problem, where a pairwise ranking loss is introduced as an optimization target~\cite{KirosSZ14_UnifyingVisualSemanticEmbeddings_arXiv}.
In the following, let $(\mathbf{x},\mathbf{y})$ denote a training pair consisting of a sheet image snippet $\mathbf{x}$ and an audio excerpt $\mathbf{y}$. As shown in~\Fig{fig:RetrievalDNN}, the network consists of two separate pathways. One pathway processes $\mathbf{x}$ and is represented by the function $f_\alpha$, where $\alpha$ are the network parameters to be trained. The other pathway, which is represented by the function $g_\beta$ with parameters $\beta$, is responsible for $\mathbf{y}$. The two functions map $\mathbf{x}$ and $\mathbf{y}$, respectively, to a $k$-dimensional vector,
where $k\in\N$ denotes the embedding dimension. To define the loss function, we need a \emph{scoring function} $s:\R^k\times\R^k\to\R$ to measure similarity in the embedding space. In our scenario, $s$ is chosen to be the cosine measure (\iec the cosine of the angle between two vectors). Furthermore, for each given training pair $(\mathbf{x},\mathbf{y})$, we assume that there are $L\in\N$ additional \emph{contrasting} examples $\mathbf{y}_\ell$ for $\ell\in\{1,2,\ldots L\}$. Then, the pairwise ranking loss (also known as \emph{max-margin hinge loss}, see~\cite{KirosSZ14_UnifyingVisualSemanticEmbeddings_arXiv}) is defined as follows:
\begin{equation}
\mathcal{L}_{\mathrm{rank}}=\sum_{(\mathbf{x},\mathbf{y})} \sum_{\ell=1}^L \max \big\{0, \gamma - s(f_\alpha(\mathbf{x}), g_\beta(\mathbf{y})) + s(f_\alpha(\mathbf{x}), g_\beta(\mathbf{y}_\ell))\big\}.
\label{eq:contrastive}
\end{equation}
In this formula, the first sum is taken over a set of training pairs $(\mathbf{x},\mathbf{y})$ (a \emph{training batch}), where each such pair comes with a separate set of contrasting examples (in practice all remaining audio samples of the current training batch). The purpose of this loss function is to encourage an embedding where the distance between matching samples $(\mathbf{x},\mathbf{y})$ is lower than the distance between mismatching samples $(\mathbf{x},\mathbf{y_\ell})$.
The parameter  $\gamma\in\R_+$ is the margin parameter of the hinge loss and, in combination with the maximum function, imposes a penalty on poorly embedded training pairs. More precisely, if the elements of a matching pair $(\mathbf{x},\mathbf{y})$ are already close in the learned embedding space and, in addition, the elements of the mismatching pairs $(\mathbf{x},\mathbf{y_\ell})$ are embedded far enough apart, the second term in the $\max$-operator goes below zero
and the respective pairs do not contribute to the overall loss. On the contrary, if the embedded elements of a matching pair are still far apart, the second term is usually above zero and will yield a substantial contribution to the overall loss.
In the training stage, the pairwise ranking loss in Equation~(\ref{eq:contrastive})
is minimized via stochastic gradient descent with respect to the network parameters $\alpha$ and $\beta$.
Once the networks represented by the functions $f_\alpha$ and $f_\beta$ are learned, the elements of matching pairs are close in the embedding space, while those of contrasting pairs are far apart (in the ideal case).
For further details concerning the network topology and the training procedure, we refer to~\cite{DorferAW17_ScoreIdentification_ISMIR,DorferSVKW_CCAProjections_IJMIR}.

Given this learned embedding space, cross-modal retrieval can be performed based on a \emph{retrieval-by-embedding paradigm}, see \Fig{fig:RetrievalDNN}. It is important to note that although the network pathways are trained simultaneously on pairs of sheet music snippets and audio excerpts, both modalities are required only at training time. At application time the two network pathways operate independently from each other. This has huge benefits in view of the cross-modal retrieval applications discussed in the previous sections. For example, in sheet-music identification and score following applications, one can first compute an embedding of an entire collection of sheet music snippets using the image embedding function $f_\alpha$. The resulting $k$-dimensional embedding vectors can be further processed and stored using suitable index structures that allow for an efficient neighborhood search. Then, given an audio excerpt as a query, the search can be performed by first projecting the query into the joint embedding space using the audio embedding function $g_\beta$ of the network, and then performing a nearest neighborhood search.

The experiments reported in~\cite{DorferAW17_ScoreIdentification_ISMIR}, which are based on 26 classical piano pieces (including the composers Bach, Haydn, Beethoven, and Chopin) and roughly 20,000 training pairs demonstrate that the end-to-end learning approach yields reasonable retrieval results for sheet music of medium complexity (\egc piano scores) and synthesized audio (used for evaluation to establish the ground truth).
In particular, combining retrieval based on snippets/excerpts with a subsequent majority voting step, the approach is capable of correctly relating sheet music and audio recordings on the piece level with high accuracy.
However, on the level of sheet music snippets (consisting of one or two bars) and audio excerpt (lasting a couple of seconds), the proposed system is not yet competitive with engineered approaches that exploit musical knowledge or are based on symbolic representations (see the approaches presented in the two previous sections).

At this stage, one may draw the conclusion that, even when comparatively little training data is available, it is still possible to use deep learning models by designing appropriate (task-specific) data augmentation strategies. First experiments showed that, when trained on only one composer, the model started to generalize to unseen scores by other composers. Therefore, we may expect that the described model will develop its full potential when provided a comprehensive dataset that consists of millions of training pairs comprising different editions and layouts of sheet music and different recorded performances.

\section*{Applications and Future Directions}
\label{sec:appl}

In this paper, we have introduced different approaches for cross-modal music retrieval aiming to bridge the gap between various music representations. Despite the remaining challenges, current technology enables a variety of music navigation and browsing applications of educational and commercial relevance.
For example, in the context of modern digital music libraries, cross-modal retrieval strategies have become an important component for content-based analysis, synchronization, indexing, and navigation in heterogeneous music collections~\cite{DammFTCKM12_DML_IJDL}. Other cross-modal applications are often subsumed under the umbrella of \emph{score following}, where the computer ``listens'' to a live performance and tries to ``read along'' in the sheet music. The output of a score following algorithm can be used for highlighting the current measure in a digital score, automatic page turning\footnote{A page turner is a person with the task of turning sheet music pages for a soloist during a performance.}, or automatic accompaniment (see, \egc~\cite{DannenbergR06_alignment_ACM}).

In the following, we describe one specific example of a prototype system to give a concrete impression of what is already possible. The \emph{Piano Music Companion} is a versatile system focused on piano music, intended to be useful for both pianists and music lovers (see~\cite{ArztBFFGLW14_MusicCompanion_ECAI}).
The system is able to identify, follow, and synchronize live performances of classical piano music, in real time.
The Piano Music Companion is a permanent listener. Whenever the pianist starts playing (regardless of which piece, or where within the piece), the companion identifies the piece, the position within the score, and continues to follow along.
This allows triggering various actions synchronized to the performed music---for instance, the current position in the sheet music is highlighted. While this is helpful for the performer and listener, further information about important themes, musical structures, and chords can be provided. In a concert setting, the system may also give hints to the listener about what to focus on at specific moments. The system may also give additional background information on the piece or composer, while telling the user where to acquire (additional) recordings of the current or related pieces.

Technically, the Piano Music Companion is based on two main components that run in parallel. The first is responsible for identifying the piece being played. To this end, symbolic fingerprinting as described above is used to continuously match the most recently detected notes of the live performance to a database of symbolically encoded sheet music (see \Fig{fig:RetrievalFingerPrinting}). Currently, the database includes the complete solo piano works by Chopin and the complete Beethoven piano sonatas, and consists of roughly 1,000,000 notes in total (about 330 pieces).
Once the piece and the rough position within the sheet music representation has been identified, the actual score following is conducted using a separate chroma-based tracking procedure, which is realized as an online variant of the matching procedure shown in Figure~\ref{fig:RetrievalChroma}.
In this way, the system combines the strengths of the respective components. The fingerprinting component is flexible, it works \emph{globally} across different pieces, and it scales over large datasets. However, since the fingerprinter's transcription step is in general faulty, the component often leads to outliers and local misalignments. This weakness is compensated by the separate chroma-based tracking component, which is less efficient but introduces a high degree of robustness (due to the chroma features). This second component is applied only \emph{locally} for tracking the score once the piece and the rough position are known.

By combining these two components, the Piano Music Companion continuously re-evaluates its hypothesis and tries to match the current input stream to the complete database. Thus, even if the musician suddenly jumps to a different score position or starts playing a completely different piece, the system is able to follow as long as the piece is part of the database. The Piano Music Companion is also highly tolerant to deviations from the notated score (due to performance errors, transcription errors, or intentional variations), and to tempo changes.  A video demonstration of our system can be found at \url{https://www.youtube.com/watch?v=SUBtND_MJZs}.

Our vision is to extend this scenario towards a \emph{Complete Classical Music Companion}. Such a system will be at one's fingertips anytime and anywhere, possibly as an application on a mobile device. Whatever source of music---be it a live concert, a DVD, a video stream, or a radio program---whatever piece of classical music, whatever instrumentation, and whoever the performers are, the companion will detect what it is listening to, inform about the music, the historical context of the piece, famous interpretations, etc., thus guiding the user in the listening process.

\begin{figure}[t]
  \centering
  \includegraphics[width=11cm]{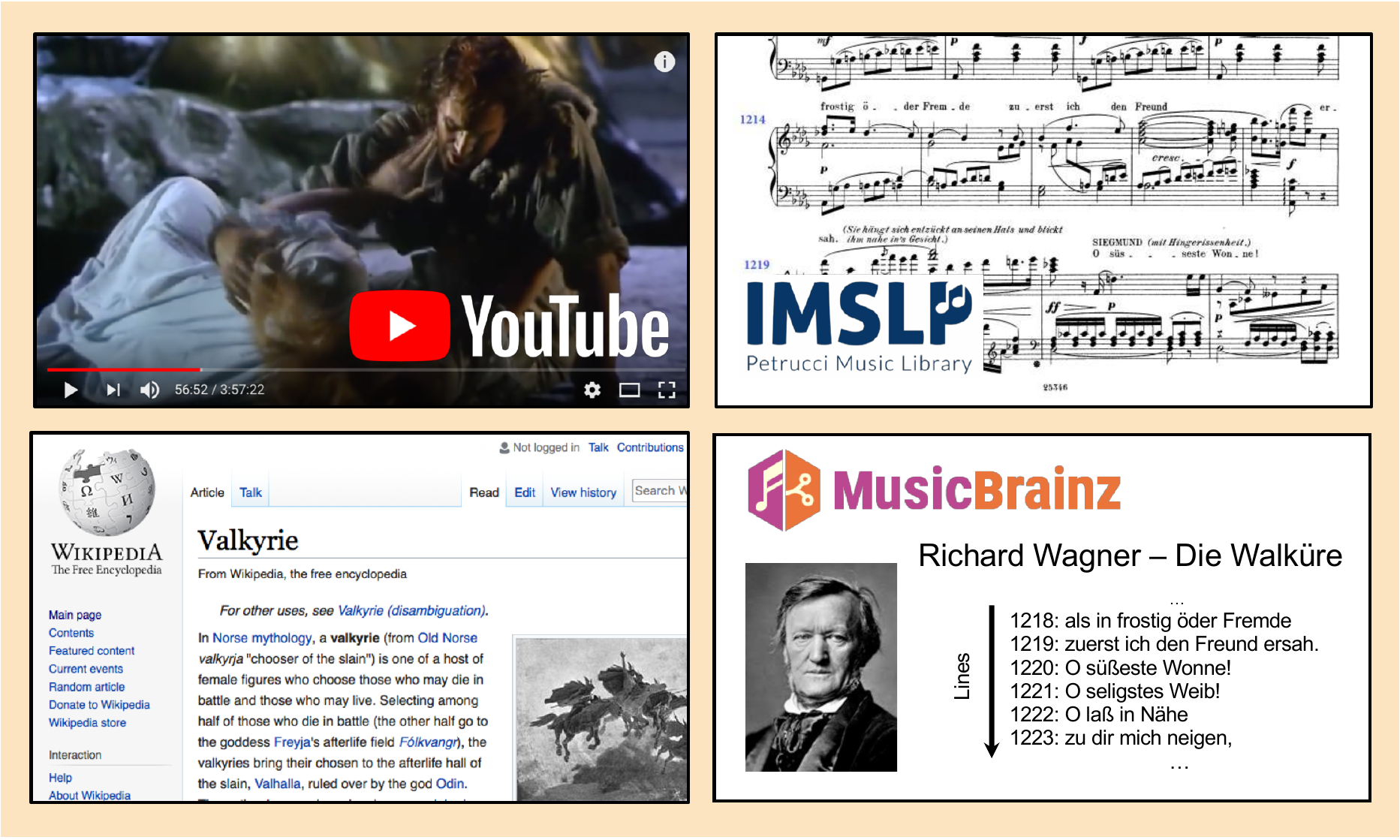}
  \caption{\small Some sources of freely accessible music data distributed over the internet. (Video image courtesy of the Warner Music Group.)}
\label{fig:AppFuture}
\end{figure}

Beyond this specific music companion scenario, cross-modal music processing techniques are essential for organizing and searching information distributed over the internet (see \Fig{fig:AppFuture}). For example, there are millions of digitized pages of sheet music publicly available on sites such as the Petrucci Music Library (IMSLP)\footnote{\url{http://imslp.org/}}. On the audio side, widely accessible music and video platforms such as YouTube offer a vast and rapidly growing corpus of music recordings. Furthermore, music-related websites as available at Wikipedia contain information of various types including text, score, images, and audio. Finally, community-driven encyclopedias such as MusicBrainz\footnote{\url{https://musicbrainz.org/}} collect and provide music-related metadata in a systematic fashion.
For example, structured websites can be used to automatically derive text-, score-, and audio-based queries to look for other musically related documents on the Web~\cite{BalkePM15_MatchingMusicalThemes_ICASSP,GasserAGGW15_ClassicalMusicWeb_ISMIR}.
Furthermore, YouTube videos may be automatically enriched with manually or automatically generated musical annotations, as recently demonstrated in~\cite{BalkeDAFPM18_JazzYoutube_Frontiers}.

This rich application potential, demonstrated in concrete application scenarios, makes cross-modal music retrieval a very active research field, which also drives research on other music processing tasks. For instance, one key challenge is to improve transformation techniques such as OMR and AMT, which are a bottleneck in many of the current approaches. Also, deep neural networks that directly learn to relate different data modalities are a very promising alternative that is getting a lot of attention now. We hope that these prospects will serve as an inspiration for the signal processing community to pay even more attention to music as a promising (and beautiful) object of study.

\section*{Acknowledgments}
\label{sec:ack}

The International Audio Laboratories Erlangen are a joint institution of the Friedrich-Alexander-Universit{\"a}t
Erlangen-N{\"u}rnberg (FAU) and the Fraunhofer-Institut f{\"u}r Integrierte Schaltungen IIS.
Meinard M{\"u}ller and Stefan Balke are supported by the German Research Foundation (DFG MU 2686/11-1).
Matthias Dorfer acknowledges financial support for early versions of this work by the Austrian Ministries Federal Ministry of Transport, Innovation, and Technology and the Federal Ministry of Science, Research, and Economy and the Province of Upper Austria via the Competence Centers for Excellent Technologies, Software Competence Center Hagenberg. The work of Andreas Arzt and Gerhard Widmer was supported by the European Research Council (ERC)
under the European Union's Horizon 2020 Framework Programme (H2020, 2014-2020) / ERC Advanced Grant Agreement n.670035, project ``Con Espressione''.

\small

\bibliographystyle{IEEEtran}

\begin{thebibliography}{10}
\providecommand{\url}[1]{#1}
\csname url@samestyle\endcsname
\providecommand{\newblock}{\relax}
\providecommand{\bibinfo}[2]{#2}
\providecommand{\BIBentrySTDinterwordspacing}{\spaceskip=0pt\relax}
\providecommand{\BIBentryALTinterwordstretchfactor}{4}
\providecommand{\BIBentryALTinterwordspacing}{\spaceskip=\fontdimen2\font plus
\BIBentryALTinterwordstretchfactor\fontdimen3\font minus
  \fontdimen4\font\relax}
\providecommand{\BIBforeignlanguage}[2]{{%
\expandafter\ifx\csname l@#1\endcsname\relax
\typeout{** WARNING: IEEEtran.bst: No hyphenation pattern has been}%
\typeout{** loaded for the language `#1'. Using the pattern for}%
\typeout{** the default language instead.}%
\else
\language=\csname l@#1\endcsname
\fi
#2}}
\providecommand{\BIBdecl}{\relax}
\BIBdecl

\bibitem{Wang03_Shazam_ISMIR}
A.~Wang, ``An industrial strength audio search algorithm,'' in
  \emph{Proceedings of the International Society for Music Information
  Retrieval Conference ({ISMIR})}, Baltimore, Maryland, USA, 2003, pp. 7--13.

\bibitem{KurthMFCC07_AutomatedSynchronization_ISMIR}
F.~Kurth, M.~M{\"u}ller, C.~Fremerey, Y.~Chang, and M.~Clausen, ``Automated
  synchronization of scanned sheet music with audio recordings,'' in
  \emph{Proceedings of the International Conference on Music Information
  Retrieval ({ISMIR})}, Vienna, Austria, Sep. 2007, pp. 261--266.

\bibitem{Mueller15_FMP_SPRINGER}
M.~M{\"u}ller, \emph{Fundamentals of Music Processing}.\hskip 1em plus 0.5em
  minus 0.4em\relax Springer Verlag, 2015.

\bibitem{CanoBKH05_FingerprintingReview_VLSI}
\BIBentryALTinterwordspacing
P.~Cano, E.~Batlle, T.~Kalker, and J.~Haitsma, ``A review of audio
  fingerprinting,'' \emph{The Journal of {VLSI} Signal Processing}, vol.~41,
  no.~3, pp. 271--284, Nov. 2005. [Online]. Available:
  \url{http://dx.doi.org/10.1007/s11265-005-4151-3}
\BIBentrySTDinterwordspacing

\bibitem{ArztBW12_SymbolicFingerprint_ISMIR}
A.~Arzt, S.~B{\"o}ck, and G.~Widmer, ``Fast identification of piece and score
  position via symbolic fingerprinting,'' in \emph{Proceedings of the
  International Society for Music Information Retrieval Conference ({ISMIR})},
  Porto, Portugal, 2012, pp. 433--438.

\bibitem{ArztWS14_TempoTranspInvariantIdent_ISMIR}
A.~Arzt, G.~Widmer, and R.~Sonnleitner, ``Tem\-po- and transposition-invariant
  identification of piece and score position,'' in \emph{Proceedings of the
  International Society for Music Information Retrieval Conference ({ISMIR})},
  Taipei, Taiwan, 2014, pp. 549--554.

\bibitem{DorferAW17_ScoreIdentification_ISMIR}
M.~Dorfer, A.~Arzt, and G.~Widmer, ``Learning audio-sheet music correspondences
  for score identification and offline alignment,'' in \emph{Proceedings of the
  International Society for Music Information Retrieval Conference ({ISMIR})},
  Suzhou, China, 2017, pp. 115--122.

\bibitem{Byrd2015_OMR_JNMR}
D.~Byrd and J.~G. Simonsen, ``Towards a standard testbed for optical music
  recognition: Definitions, metrics, and page images,'' \emph{Journal of New
  Music Research}, vol.~44, no.~3, pp. 169--195, 2015.

\bibitem{BenetosDGKK13_MusicTranscription_JIIS}
\BIBentryALTinterwordspacing
E.~Benetos, S.~Dixon, D.~Giannoulis, H.~Kirchhoff, and A.~Klapuri, ``Automatic
  music transcription: challenges and future directions,'' \emph{Journal of
  Intelligent Information Systems}, vol.~41, no.~3, pp. 407--434, 2013.
  [Online]. Available: \url{http://dx.doi.org/10.1007/s10844-013-0258-3}
\BIBentrySTDinterwordspacing

\bibitem{Gomez06_PhD}
E.~G{\'o}mez, ``Tonal description of music audio signals,'' {P}h{D} Thesis,
  Universitat Pompeu Fabra, Barcelona, Spain, 2006.

\bibitem{Mueller07_InformationRetrieval_SPRINGER}
M.~M{\"u}ller, \emph{Information Retrieval for Music and Motion}.\hskip 1em
  plus 0.5em minus 0.4em\relax Springer Verlag, 2007.

\bibitem{BarlowM75_MusicalThemes_BOOK}
H.~Barlow and S.~Morgenstern, \emph{A Dictionary of Musical Themes}, revised
  edition third printing~ed.\hskip 1em plus 0.5em minus 0.4em\relax Crown
  Publishers, Inc., 1975.

\bibitem{BalkePM15_MatchingMusicalThemes_ICASSP}
S.~Balke, S.~P. Achankunju, and M.~M{\"u}ller, ``Matching musical themes based
  on noisy {OCR} and {OMR} input,'' in \emph{Proceedings of the {IEEE}
  International Conference on Acoustics, Speech, and Signal Processing
  ({ICASSP})}, Brisbane, Australia, 2015, pp. 703--707.

\bibitem{BalkeALM16_BarlowRetrieval_ICASSP}
S.~Balke, V.~Arifi-M{\"u}ller, L.~Lamprecht, and M.~M{\"u}ller, ``Retrieving
  audio recordings using musical themes,'' in \emph{Proceedings of the {IEEE}
  International Conference on Acoustics, Speech, and Signal Processing
  ({ICASSP})}, Shanghai, China, 2016, pp. 281--285.

\bibitem{SalamonSG13_Retrieval_IJMRI}
\BIBentryALTinterwordspacing
J.~Salamon, J.~Serr{\`{a}}, and E.~G{\'{o}}mez, ``Tonal representations for
  music retrieval: from version identification to query-by-humming,''
  \emph{International Journal of Multimedia Information Retrieval}, vol.~2,
  no.~1, pp. 45--58, 2013. [Online]. Available:
  \url{http://dx.doi.org/10.1007/s13735-012-0026-0}
\BIBentrySTDinterwordspacing

\bibitem{BoschBSG16_MelodyExtraction_ISMIR}
J.~S. Juan J.~Bosch, Rachel M.~Bittner and E.~G{\'o}mez, ``A comparison of
  melody extraction methods based on source-filter modelling,'' in
  \emph{Proceedings of the International Conference on Music Information
  Retrieval ({ISMIR})}, New York City, USA, 2016, pp. 571--577.

\bibitem{CaseyRS08_MinimumDistances_IEEE-TASLP}
M.~A. Casey, C.~Rhodes, and M.~Slaney, ``Analysis of minimum distances in
  high-dimensional musical spaces,'' \emph{{IEEE} Transactions on Audio,
  Speech, and Language Processing}, vol.~16, no.~5, pp. 1015--1028, 2008.

\bibitem{GroscheM12_RetrievalShingles_ICASSP}
P.~Grosche and M.~M{\"u}ller, ``Toward characteristic audio shingles for
  efficient cross-version music retrieval,'' in \emph{Proceedings of the {IEEE}
  International Conference on Acoustics, Speech, and Signal Processing
  ({ICASSP})}, Kyoto, Japan, 2012, pp. 473--476.

\bibitem{SonnleitnerW16_QuadFingerp_TASLP}
R.~Sonnleitner and G.~Widmer, ``Robust quad-based audio fingerprinting,''
  \emph{IEEE Transactions on Audio, Speech, and Language Processing}, vol.~24,
  no.~3, pp. 409--421, 2016.

\bibitem{SixL14_PanakoAcousFP_ISMIR}
J.~Six and M.~Leman, ``Panako -- {A} scalable acoustic fingerprinting system
  handling time-scale and pitch modification,'' in \emph{Proceedings of the
  International Conference on Music Information Retrieval ({ISMIR})}, Taipei,
  Taiwan, 2014, pp. 259--264.

\bibitem{SerraGH10_coversong_BOOKCHAP}
J.~Serr{\`a}, E.~G{\'o}mez, and P.~Herrera, ``Audio cover song identification
  and similarity: Background, approaches, evaluation and beyond,'' in
  \emph{Advances in Music Information Retrieval}, ser. Studies in Computational
  Intelligence, Z.~W. Ras and A.~A. Wieczorkowska, Eds.\hskip 1em plus 0.5em
  minus 0.4em\relax Berlin, Germany: Springer, 2010, vol. 274, ch.~14, pp.
  307--332.

\bibitem{BoeckS12_TranscriptionRecurrentNetwork_ICASSP}
S.~B{\"{o}}ck and M.~Schedl, ``Polyphonic piano note transcription with
  recurrent neural networks,'' in \emph{{IEEE} International Conference on
  Acoustics, Speech and Signal Processing ({ICASSP})}, Kyoto, Japan, March
  2012, pp. 121--124.

\bibitem{SigtiaBD16_DNNPolyPianoTrans_TASLP}
S.~Sigtia, E.~Benetos, and S.~Dixon, ``An end-to-end neural network for
  polyphonic piano music transcription,'' \emph{{IEEE/ACM} Transactions on
  Audio, Speech, and Language Processing}, vol.~24, no.~5, pp. 927--939, 2016.

\bibitem{Schmidhuber15_DeepLearningOverview_NN}
J.~Schmidhuber, ``Deep learning in neural networks: An overview,'' \emph{Neural
  Networks}, vol.~61, pp. 85--117, 2015.

\bibitem{KirosSZ14_UnifyingVisualSemanticEmbeddings_arXiv}
R.~Kiros, R.~Salakhutdinov, and R.~S. Zemel, ``Unifying visual-semantic
  embeddings with multimodal neural language models,'' \emph{arXiv preprint
  (arXiv:1411.2539)}, 2014.

\bibitem{DorferSVKW_CCAProjections_IJMIR}
M.~Dorfer, J.~Schl{\"u}ter, A.~Vall, F.~Korzeniowski, and G.~Widmer,
  ``End-to-end cross-modality retrieval with cca projections and pairwise
  ranking loss,'' \emph{International Journal of Multimedia Information
  Retrieval}, 2017.

\bibitem{DammFTCKM12_DML_IJDL}
D.~Damm, C.~Fremerey, V.~Thomas, M.~Clausen, F.~Kurth, and M.~M{\"u}ller, ``A
  digital library framework for heterogeneous music collections: from document
  acquisition to cross-modal interaction,'' \emph{International Journal on
  Digital Libraries: Special Issue on Music Digital Libraries}, vol.~12, no.
  2-3, pp. 53--71, 2012.

\bibitem{DannenbergR06_alignment_ACM}
R.~B. Dannenberg and C.~Raphael, ``Music score alignment and computer
  accompaniment,'' \emph{Communications of the {ACM}, Special Issue: Music
  Information Retrieval}, vol.~49, no.~8, pp. 38--43, 2006.

\bibitem{ArztBFFGLW14_MusicCompanion_ECAI}
A.~Arzt, S.~B\"{o}ck, S.~Flossmann, H.~Frostel, M.~Gasser, C.~C.~S. Liem, and
  G.~Widmer, ``The piano music companion,'' in \emph{Proceedings of the
  European Conference on Artificial Intelligence}, Prague, Czech Republic,
  2014, pp. 1221--1222.

\bibitem{GasserAGGW15_ClassicalMusicWeb_ISMIR}
M.~Gasser, A.~Arzt, T.~Gadermaier, M.~Grachten, and G.~Widmer, ``Classical
  music on the web - user interfaces and data representations,'' in
  \emph{Proceedings of the International Conference on Music Information
  Retrieval ({ISMIR})}, 2015, pp. 571--577.

\bibitem{BalkeDAFPM18_JazzYoutube_Frontiers}
S.~Balke, C.~Dittmar, J.~Abe{\ss}er, K.~Frieler, M.~Pfleiderer, and
  M.~M{\"u}ller, ``Bridging the {G}ap: {E}nriching {Y}ou{T}ube videos with jazz
  music annotations,'' \emph{Frontiers in Digital Humanities}, vol.~5, 2018.

\end{thebibliography}


\appendix

\linespread{\appendixlinespacing}
\selectfont
\normalsize

\section*{Bibliography of Authors}
\label{sec:authors}

\subsection*{Meinard Müller (meinard.mueller@audiolabs-erlangen.de)}

Meinard Müller received the Diploma degree in mathematics and the Ph.D. degree in computer science from the University of Bonn, Germany, in 1997 and 2001, respectively. Since 2012, he holds a professorship for Semantic Audio Signal Processing at the International Audio Laboratories Erlangen. His recent research interests include music processing, music information retrieval, and audio signal processing. Meinard Müller has coauthored more than 100 peer-reviewed scientific papers, wrote a monograph titled ``Information Retrieval for Music and Motion'' (Springer, 2007) as well as a textbook titled ``Fundamentals of Music Processing'' (Springer, 2015, www.music-processing.de).

\subsection*{Andreas Arzt (andreas.arzt@jku.at)}

Andreas Arzt is an assistant professor at the Johannes Kepler University Linz. He studied computer science at the University of Vienna and at the Vienna University of Technology, in 2006 and 2008, respectively. In 2016, he finished his PhD thesis with the topic ``Flexible and Robust Music Tracking'' at the Department of Computational Perception at the Johannes Kepler University in Linz, Austria. His research interests include real-time music tracking, music synchronization and music identification.

\subsection*{Stefan Balke (stefan.balke@audiolabs-erlangen.de)}

Stefan Balke studied electrical engineering at the Leibniz Universität Hannover, Germany, in 2013. In early 2018, he completed his PhD in the Semantic Audio Signal Processing Group headed by Prof. Meinard Müller at the International Audio Laboratories Erlangen. Afterwards, he joined the Institute of Computational Perception led by Gerhard Widmer at the Johannes Kepler University Linz. His research interests include music information retrieval, machine learning, and multimedia retrieval. In his spare time, he plays trumpet in several local bands and projects.

\subsection*{Matthias Dorfer (matthias.dorfer@jku.at)}

Matthias Dorfer studied Medical Informatics at Vienna University of Technology with a focus on computational image analysis and machine learning. After finishing his Master studies he worked two years as an industrial researcher in the field of medical image analysis. Since April 2015 he is a PhD student at the Department of Computational Perception at Johannes Kepler University Linz under the supervision of Professor Gerhard Widmer. His research interests are artificial neural networks, especially multimodality deep learning, and audio-visual representation learning.

\subsection*{Gerhard Widmer (gerhard.widmer@jku.at)}

Gerhard Widmer received his Ph.D. degree in computer science in 1989 from Technische Univesität Wien, Austria. He is a professor and head of the Department of Computational Perception at Johannes Kepler University, Linz, Austria, and head of the Intelligent Music Processing and Machine Learning Group at the Austrian Research Institute for Artificial Intelligence (OFAI), Vienna, Austria. His research interests include AI, machine learning, and intelligent music processing. He is a Fellow of the European Association for Artificial Intelligence (EurAI), has been awarded Austrias highest research awards, the START Prize (1998), the Wittgenstein Award (2009), and currently holds an ERC Advanced Grant for research on computational models of expressivity in music.
%
%

\end{document}